\documentstyle[preprint,aps,prb,epsf,floats]{revtex}
 
\newcommand{\la}[1]{\label{#1}}

\newcommand{\be}{\begin{equation}}
\newcommand{\ee}{\end{equation}}
\newcommand{\ba}{\begin{eqnarray}}
\newcommand{\ea}{\end{eqnarray}}
\newcommand{\bi}{\begin{itemize}}
\newcommand{\ei}{\end{itemize}}

\newcommand{\nr}[1]{(\ref{#1})}
\newcommand{\bfx}{\mbox{\bf x}}
\newcommand{\fr}[2]{{\frac{#1}{#2}}}
\newcommand{\msbar}{\overline{\mbox{\rm MS}}}

\begin{document}

\draft
\title{Masses and Phase Structure in the Ginzburg-Landau Model}
\author{K. Kajantie$^{1,2}$\footnote{keijo.kajantie@cern.ch},
M. Karjalainen$^{2}$\footnote{mikarjal@phcu.helsinki.fi},
M. Laine$^{3}$\footnote{m.laine@thphys.uni-heidelberg.de} and
J. Peisa$^{4}$\footnote{peisa@amtp.liv.ac.uk}
}
\address{$^{1}$Theory Division, CERN, CH-1211 Geneva 23,
Switzerland}
\address{$^{2}$Department of Physics,
P.O.Box 9, 00014 University of Helsinki, Finland}
\address{$^{3}$Institut f\"ur Theoretische Physik,
Philosophenweg 16, D-69120 Heidelberg, Germany}
\address{$^{4}$Department of Mathematical Sciences, 
University of Liverpool, Liverpool L69 3BX, UK}
\date{August, 1997}
\maketitle

\vspace*{-8.0cm}
\noindent
\hspace*{6.5cm} 
\mbox{CERN-TH/97-62, HD-THEP-97-05, cond-mat/9704056}
\vspace*{6.8cm}

\begin{abstract}
We study numerically the phase structure of the Ginzburg-Landau 
model, with particular emphasis on mass measurements.
There is no local gauge invariant order parameter, 
but we find that there is a phase transition
characterized by a vanishing photon mass.
For type I superconductors the transition is of 1st order. 
For type II 1st order is excluded by susceptibility analysis,
but the photon correlation length suggests 2nd order
critical behaviour with $\nu\sim\fr12$. 
The scalar mass, in contrast, does
not show clear critical behaviour in the type II regime 
for $V\to\infty$, contrary to the conventional picture. 
\end{abstract}

\section{Introduction}

One of the most interesting phase transitions known is that 
some materials become superconductive at low temperatures.
While the microscopic dynamics behind the phenomenon is 
complicated, there exists a 
simple effective theory for describing this transition. The
effective theory is just the Ginzburg-Landau (G-L) model, 
or a three-dimensional U(1) + complex scalar gauge theory.
The modulus squared of the scalar field represents 
the density of superconductive electrons.
The G-L model might have other applications as well, 
such as the nematic-to-smectic-A transition
in liquid crystals~\cite{lc}.

There exists an extensive literature on the phase diagram of the
G-L model\cite{hlm,dh,b,kleinert,hk,arg}. 
Defining the standard G-L parameter as $x=m_H^2/2m_W^2$ 
(where $m_H$ is the inverse scalar correlation length, or coherence
length, and $m_W$ is the inverse vector correlation length, 
or penetration depth), the conventional
picture is that the transition is of first order
for small $x$ (type I superconductors), gets weaker with increasing $x$,
has a tricritical point at some $x_c$, and remains of second order for 
$x>x_c$ (type II superconductors)
~\cite{othertheories}. 
This picture is based on perturbative, renormalization group 
and lattice studies, often in a dual theory.
These studies have nevertheless
not been conclusive and there
have been arguments concerning, for instance, the universality
class of the provisional second order transition in the type II 
regime ($x > 1/2$)~\cite{arg}. 
 
The purpose of this paper is to study
the phase diagram numerically. We improve significantly upon 
earlier numerical results \cite{dh,b}, by having a much
finer lattice (smaller lattice constant $a$ in physical units) and 
by measuring, for the first time, the different correlation lengths.
While the infinite volume and continuum extrapolations are numerically 
demanding and thus the conclusions based on a series
of finite lattices can never be quite complete,  
we nevertheless find indications of 
quite an unexpected pattern in the type II regime.
We also point out directions for future investigations
of this issue.

The paper is organized as follows. 
In Sec.~\ref{model} we specify the model studied, 
in Sec.~\ref{discretization} we discuss how it is discretized, and 
in Sec.~\ref{simulations} how the simulations are
organized and what our results are.
Sec.~\ref{discussion} is a discussion.

\section{The model}
\la{model}

Let us first define the model unambiguously. It is a
locally gauge invariant 3-dimensional continuum
U(1) + complex scalar field theory defined by the functional integral
\ba
Z=&&\int {\cal D}A_i{\cal D}\phi\,\exp\bigl[-S(A_i,\phi)\bigr], \la{z} \\
S=&&\int d^3x\biggl[\fr14(\partial_iA_j-\partial_jA_i)^2+
|(\partial_i+ie_3A_i)\phi|^2 + \nonumber \\
&&\qquad + m_3^2 \phi^*\phi + \lambda_3 \left(\phi^*\phi\right)^2\bigg]. 
\label{ac}
\end{eqnarray}
The couplings $e_3^2,\lambda_3$ have the dimension of mass 
(in units $\hbar=c=1$) and, factoring out one scale ($e_3^2$),
the free energy density $f$ of the model depends
on the two massless ratios  
\be
y\equiv \frac{m_3^2}{e_3^4}, \quad x\equiv\frac{\lambda_3}{e_3^2},
\ee
so that
\be
Z=\exp\bigl[-Ve_3^6 f(y,x)\bigr].
\ee
The phase diagram of the theory can thus be drawn in the 
$(x,y)$-plane.

Since the theory in eq.~\nr{ac} is a continuum field
theory, one has to consider ultraviolet renormalization.
There is a linear 1-loop and logarithmic 2-loop 
divergence~\cite{parisi} for the mass parameter $m_3^2$. 
In the $\msbar$ dimensional regularization scheme in 
$3-2\epsilon$ dimensions, the renormalized mass parameter
becomes \cite{fkrs}
\be
m_3^2(\mu) =  {-4e_3^4+8\lambda_3e_3^2-8\lambda_3^2\over16\pi^2}
\log\frac{\Lambda_m}{\mu},
\la{m3}
\ee
where
$\Lambda_m$ specifies the theory. To be more precise, 
we thus define
$y\equiv m_3^2(e_3^2)/e_3^4$, which specifies the 
continuum theory at the full quantum level
equally well as $\Lambda_m$. 

The physical values of $x$ and $y$ depend on the microscopic theory
behind the effective theory in eq.~\nr{ac}. 
For reference, for usual superconductors
in the notation of Ref.~\cite{kleinert}, 
\be
y={1\over rq^4}\biggl({T\over T_c}-1\biggr),\quad
x={g\over (rq)^2}\sim \frac{0.01}{r^2},
\ee
with $g\sim 10^{-6},q\sim 0.01,r\le1$. 
For high $T_c$ superconductors $x$ can be
$\gg1$. At present we consider the theory \nr{ac} as
such and questions of validity, like the need to include 
$(\phi^*\phi)^3$-terms, do not enter.

The phase diagram of the G-L model on the $y,x>0$ plane
(see Fig.\ref{ycx})
contains a curve $y=y_c(x)$ along which the system 
has a first order transition for small~$x$. To one vector loop this
curve is simply given by $y_c=1/(18\pi^2x)$. We shall call the region
$y>y_c(x)$ the normal and the region $y<y_c(x)$ the superconducting
(SC) phase. At small $x$ the reliability of perturbation 
theory has been verified numerically~\cite{u1}. However, 
perturbation theory gets worse at large $x$ and the issue now is what
happens then.

\begin{figure}[t]
\hspace{1cm}
\epsfysize=8cm
\centerline{\epsffile{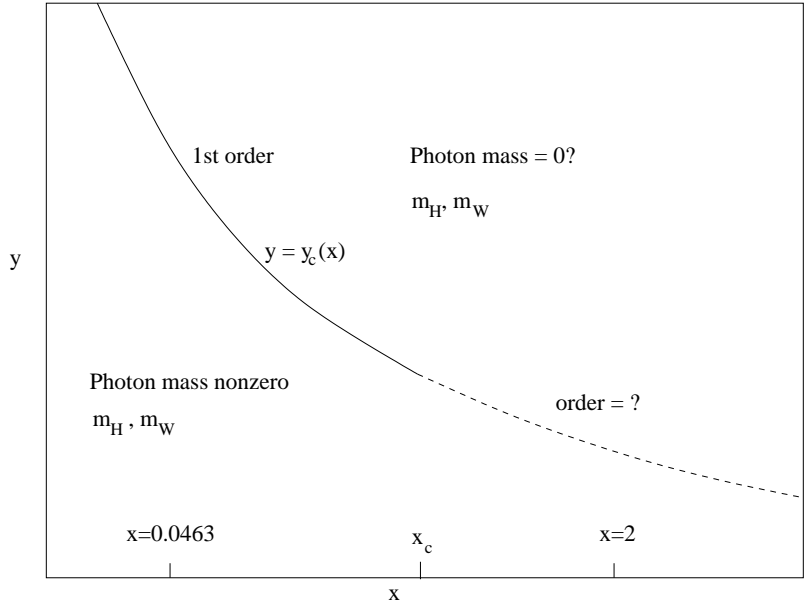}}
\caption[a]{The qualitative phase diagram of the G-L theory.}
\la{ycx}
\end{figure}

\section{Discretization}
\la{discretization}

To latticize ($a$ = lattice spacing)
the theory with the fixed continuum variables $x,y$, 
we introduce the link field $U_i(\bfx)=
\exp[iae_3A_i(\bfx)]\equiv\exp[i\alpha_i(\bfx)]$.
Relating the
counterterms in the $\msbar$ and lattice regularization schemes
\cite{laine}, the lattice action becomes~\cite{noncmpct} 
\begin{eqnarray}
S & = & \beta_G \!\!\sum_{\bfx,i<j}
\Bigl[1-\cos \hat{F}_{ij}(\bfx)\Bigr] \nonumber \\
  & & 
  -\beta_H \sum_{\bfx, i} {\mbox{Re}}\, 
\phi^*(\bfx) U_i(\bfx)\phi(\bfx+\hat{i}) \nonumber \\   
  & & +{\beta_H\over2}\sum_{\bfx} 
 \phi^*(\bfx) \phi(\bfx)\biggl[6+{y\over\beta_G^2}-
{3.1759115(1+2x)\over2\pi\beta_G} \nonumber\\
&&-{(-4+8x-8x^2)(\log6\beta_G+0.09)+25.5+4.6x\over16\pi^2\beta_G^2}
\biggr] \nonumber\\
&&+ {x\beta_H^2\over4\beta_G}
\sum_{\bfx} \left[\phi^*(\bfx)\phi(\bfx)\right]^2, \label{laq}
\end{eqnarray}
where $\beta_H$ is so far arbitrary, $\beta_G=1/e_3^2a$,  and
\be
\hat{F}_{ij}(\bfx)=\alpha_i(\bfx)+\alpha_j(\bfx+\hat i)-
\alpha_i(\bfx+\hat j)-\alpha_j(\bfx). \la{Fijlat}
\ee
We have scaled the continuum scalar field to a dimensionless lattice
field by $\phi^*\phi\to \beta_H\phi^*\phi/2a$, but further rescalings
(by a specific choice of $\beta_H$)
are possible; we thus 
scale the coefficient of  $\phi^*\phi$ to be +1.
For a given continuum theory ($e_3^2,y,x$), eqs.~\nr{laq}--\nr{Fijlat}
specify up to terms of order $e_3^2a$ the corresponding lattice action.
It should be noted that the complicated counterterm expression 
in the square brackets in eq.~\nr{laq} only affects the value of $y_c$
for given $x$ (the counterterm
guarantees that the limit $y_c(a\to 0)$ exists), 
but not the qualitative structure of the phase diagram.

One of the most essential points of the present 
lattice simulations is the extrapolation 
to the continuum limit: first the infinite volume $V\to\infty$ at fixed
lattice spacing $a$, then $a\to 0$. 
To estimate the required sizes of $V=(N a)^3$ and $a$, consider a system 
with a typical correlation length $\xi$. Then one has to satisfy 
(on a periodic lattice) $a\ll\xi\ll Na/2$ or, in physical units,
\be
e_3^2a = \frac{1}{\beta_G}\ll e_3^2\xi \ll \frac{N}{2\beta_G}.
\la{constraints}
\ee
We observe that $e_3^2\xi \sim 1$ and take
$\beta_G=4,6$ so that $1/\beta_G\ll e_3^2\xi$; and 
$N=32,\ldots,64$ so that $N/(2\beta_G) \gg e_3^2\xi$.
Note that Ref.~\cite{b} had $\beta_G=0.2$, $N\le15$, 
so that the lattice spacing $a$ was larger than 
the typical correlation lengths.

Apart from the UV-cutoff effects discussed,  
there is another effect related to a finite $a$. 
Indeed, we use a compact formulation for the U(1) gauge field, 
which changes the topology of the continuum theory and 
implies that the photon becomes massive~\cite{polyakov}.
However, a semiclassical 
computation for this Polyakov mass gives~\cite{ambjorn}
\be
\frac{m_\gamma^P}{e_3^2}
=\pi(2 \beta_G)^{3/2}\exp\biggl[-{3.176\pi\over4}\beta_G\biggr]. \la{pmass}
\ee
Thus for $\beta_G\ge 4$ this photon mass
(as well as the monopole density from which it originates)
should be negligible ($m_\gamma^P/e_3^2<0.01$) and our results the same 
as in the non-compact formulation within statistical errors.

\section{Simulations and results}
\la{simulations}

For the simulations we choose two values of $x$, 
$x=0.0463$ and $x=2$, corresponding to a strongly type I and type II
superconductor. We then measure averages of local or bilocal
gauge invariant quantities and
locate the critical curve $y=y_c(x)$ on which
the system changes its properties.
Note that there is no 
local gauge invariant order parameter which would vanish
in either of the phases.
However, we shall find that the photon mass, measured from a correlator,
vanishes in one of the phases within errors.

The phase transition is located by  finding the maximum in $y$ 
of the susceptibility $\chi$ defined by 
\be
\chi = e_3^2 V \left\langle (\overline{\phi^*\phi}
     - \langle \overline{\phi^*\phi}\rangle)^2\right\rangle, \la{susc}
\ee
where $\overline{\phi^*\phi}$ is the volume average, 
$\overline{\phi^*\phi} = V^{-1}\int\! d^3x \phi^*\phi$, 
and by studying its large-$V$ behavior~\cite{finitesize}. 
If there is a first order transition, the distribution of 
$\overline{\phi^*\phi}$ 
precisely at $y_c(x)$ has two peaks which remain at fixed distance
and get narrower when $V\to\infty$. Then the maximum of
$\chi$ grows as the volume $V$. In a second order
transition the expected behaviour is $\sim V^\kappa$, $\kappa$
being a critical exponent. If $\chi \sim V^0$, then either
$\kappa\le 0$, or the transition
is of higher than second order or absent.

The susceptibility 
maximum is plotted in Fig. \ref{suscmax}. 
One sees a very clear difference between $x=0.0463$ and $x=2$. The  
behavior of the system at $x=0.0463$ at the largest volumes
indicates a linear 1st order behavior.
At $x=2$, in contrast, the transition 
is not of first order. If the transition is of second order
the critical exponent $\kappa$ is close to zero, as noticed already
in \cite{dh,b}. However, a still higher order 
transition (or a smooth crossover) cannot
be excluded based on these measurements. We thus turn to masses.

For the mass measurements we use the scalar operator 
$\phi^{\ast}\phi$ and the two vector operators
$\phi^*D_i\phi$ and $\epsilon_{ijk}F_{jk}$. On the lattice these are
\ba
O(\bfx)&=&\phi^{\ast}(\bfx)\phi(\bfx), \\
O_i(\bfx)&=& \mathop{\rm Im}\phi^{\ast}(\bfx)U_i(\bfx)\phi(\bfx+\hat i),
\\
\tilde O_i(\bfx)&=&\epsilon_{ijk}\sin\hat F_{jk}(\bfx).
\ea

\begin{figure}[t]
\hspace{1cm}
\epsfysize=9cm
\centerline{\epsffile{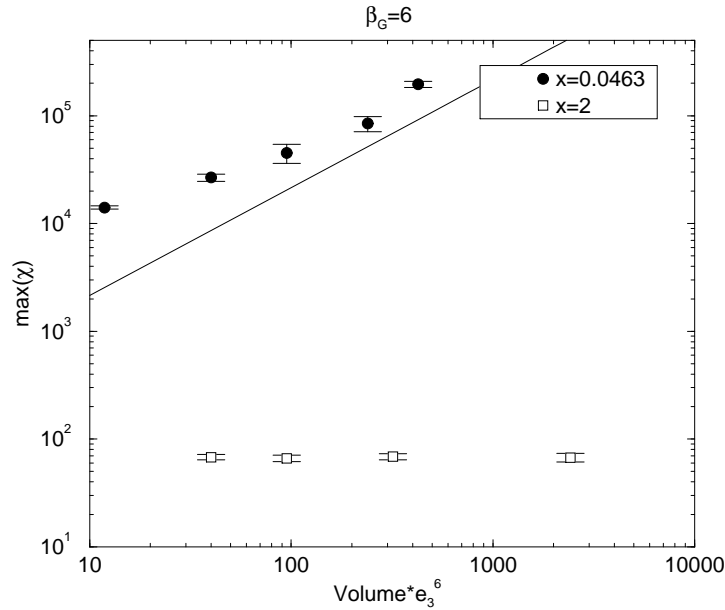}}
\caption[a]{The maximum of the 
           susceptibility $\chi$ as a function of
	   volume. The straight line is $\sim V$. }
\la{suscmax}
\end{figure}

The correlation masses are measured 
(from a lattice of size $N_x^2N_z$) 
by first summing over planes, possibly with momentum $p=2\pi/(aN_x)$,
\be
O(z;p)=\sum_{x,y}e^{ipx}O(x,y,z), \la{Ozp}
\ee
and then studying the large-$t$ behavior of
\be
G(t;p)={1\over N_x^2N_z}\sum_z 
\langle O(z;p)O^\ast(z+t;p) \rangle.
\ee
The momentum is needed for the correlator of $\tilde O_{3}$, 
used to measure the photon mass~\cite{bp} (it can also be used
for the very light scalar mass, to get a better signal in 
a finite volume): without the factor $e^{ipx}$ in eq.~\nr{Ozp}, 
the plane average for $\tilde O_3$ would simply vanish. 
In perturbation theory, 
one finds for the asymptotic behavior of this 
correlator in the continuum limit,
\be
G_{3}(t)={A_\gamma\over\beta_G}{ap^2\over2E}e^{-Et},\quad E^2=p^2+
m_\gamma^2, \la{Gt}
\ee
where $m_\gamma=0$ in the normal phase.
At 1-loop level, 
\be
A_\gamma=1-\frac{1}{24\pi\sqrt{y}}.
\ee  
Very close to the critical point $y\sim 0$ the expansion for
$A_\gamma$ thus breaks down, and the form of the correlation
function is determined by an anomalous dimension. 
For our datapoints, the functional form in eq.~\nr{Gt}
fits the data well and the energy $E$ is measured from 
the exponential fall-off.

To improve the projection to the low-lying mass states, 
it is indispensable~\cite{old} 
to use blocking techniques~\cite{phtw}
to define extended operators, and to make a 
mixing analysis~\cite{phtw} between operators at 
different blocking levels, to search for the linear
combination giving the best signal.  
We find the best results with blocking level 3 
for the scalar mass. The mixing analysis allows
to get a good signal already at a relatively small $t$, 
$t\sim (2-5) a$. We denote by $m_\gamma$ the lowest lying 
vector state, and by $m_W$ the first 
excitation (where it can be determined).

\begin{figure}[h]
\hspace{1cm}
\epsfysize=8cm
\centerline{\epsffile{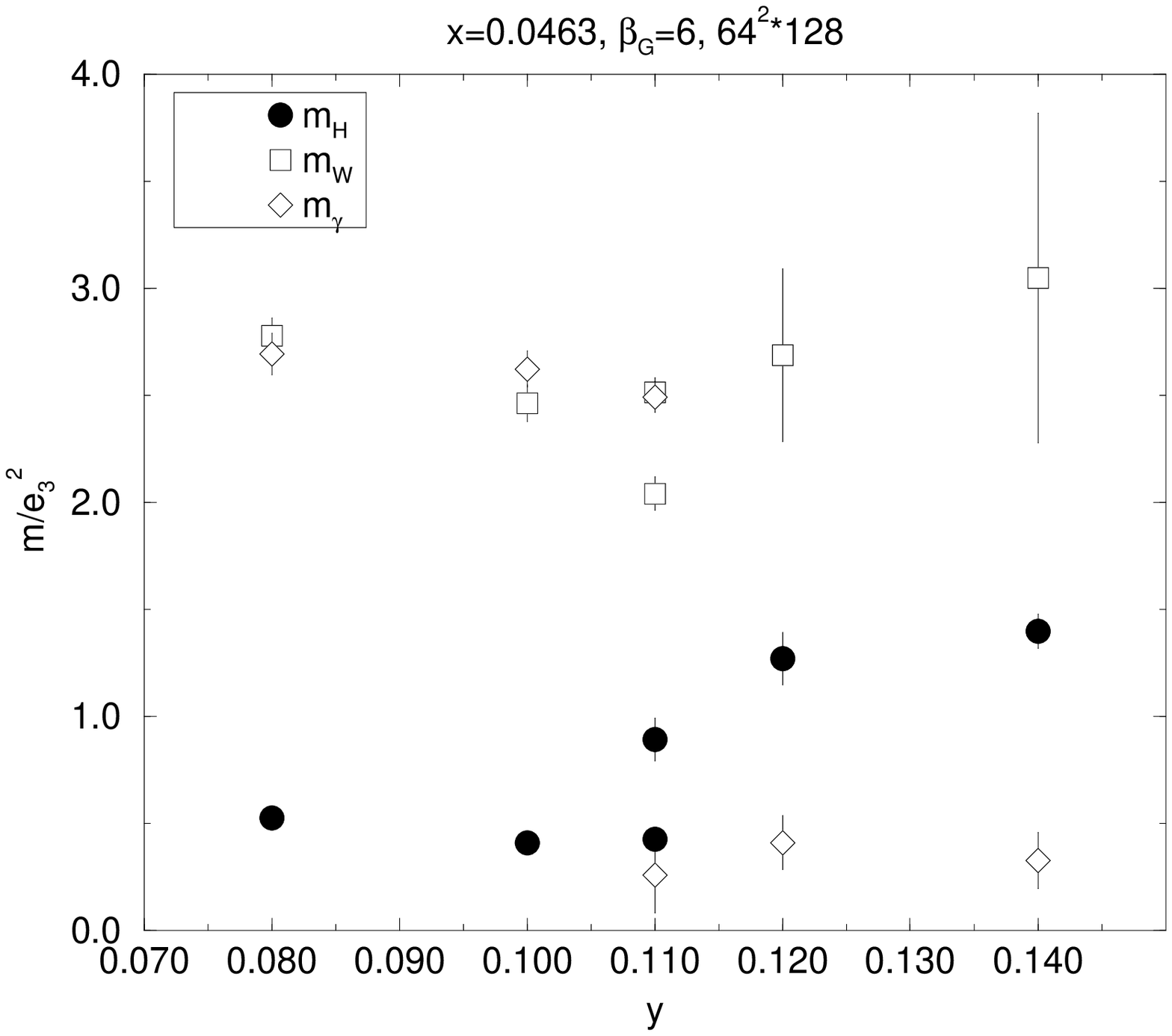}}

\vspace*{-0.5cm}

\epsfysize=8cm
\centerline{\epsffile{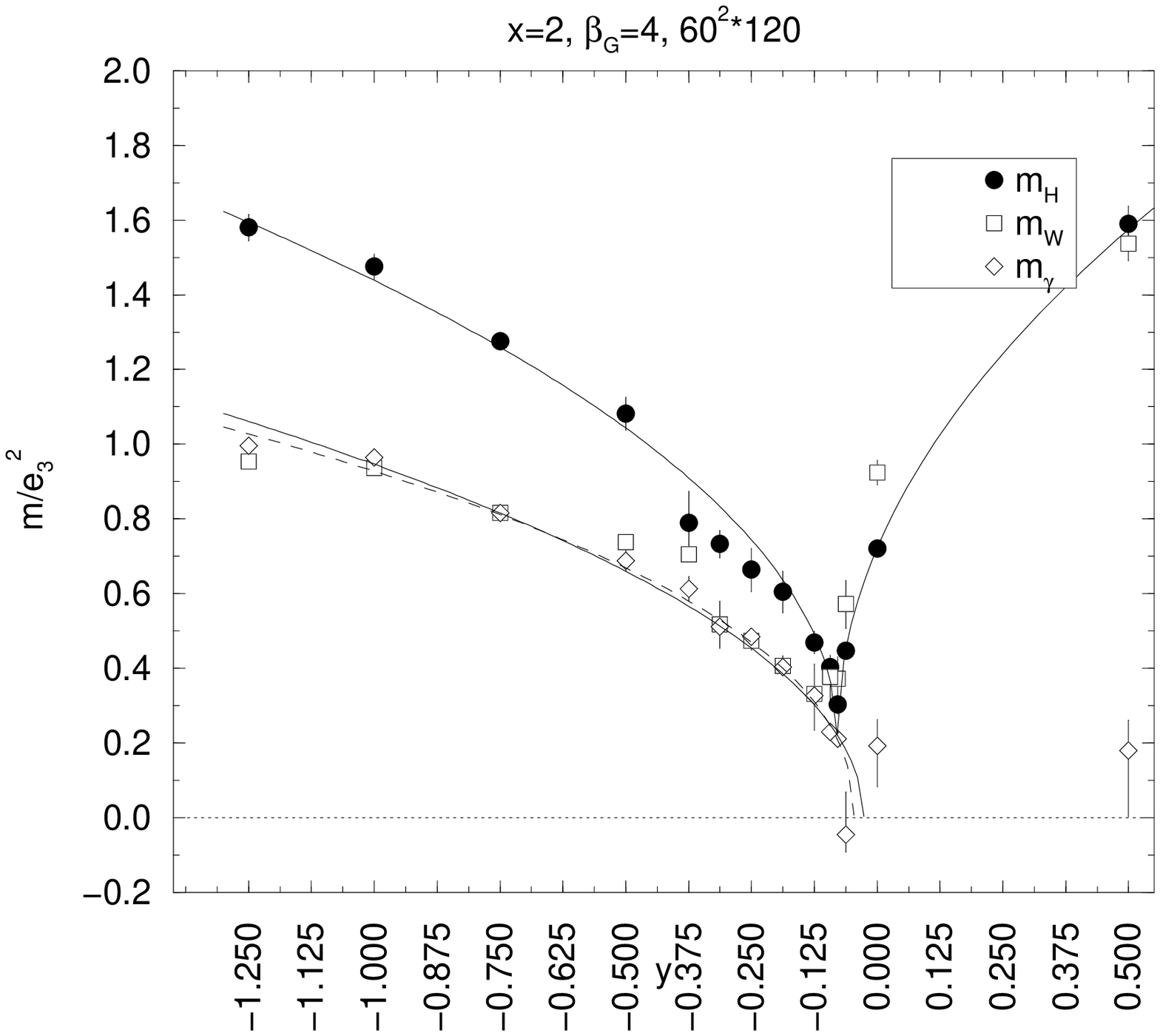}}
\caption[a]{The masses at $x=0.0463$ (top) and $x=2$ (bottom).
Note the discontinuous and continuous behaviour of $m_\gamma$ 
in the top and bottom figures, respectively. 
The values of $m_\gamma$ 
in the top figure for $y=0.12,0.14$ are 2\ldots3
standard deviations from zero, but all or most of this deviation 
is expected to be 
due to the absence of a mixing analysis at these data points.
The curves in the bottom figure represent fits $\sim A|y-y_c|^\nu$. 
The dashed curve for $m_\gamma$ is with a free 
exponent $\nu=0.44(2)$ over the whole $y$-range
whereas for the solid curves, $\nu$ has been fixed
to $\nu=\fr12$.}
\la{masses}
\end{figure}

The masses ($=1/\xi$) near the transition are shown in 
Fig.~\ref{masses}. For type I superconductors ($x=0.0463$)
one observes in the SC phase one scalar with a
rather small mass $m_H$.
The vector operators couple dominantly to 
a single state
of a larger mass $m_W=m_\gamma$. These are the standard 
(inverse) coherence length and
penetration depth. Close to the transition point one can 
observe both phases simultaneously, and the masses are
discontinuous. Above the transition 
the scalar excitation is there with a larger mass,
but the vector operators couple to two quite different states.
There is 
a $\phi^*\phi$ bound state of large mass $m_W$,
seen dominantly by the operator $O_i$, 
while the operator 
$\tilde O_3$ sees the photon of mass $m_\gamma$ which is consistent 
with zero within $2\sigma$. 
The picture here is the standard one of a first order transition
with $m_\gamma$ as an effective order parameter.

For type II superconductors ($x=2$) a 1st order transition was excluded
by the susceptibility analysis: no two-peak structure exists. However, 
Fig.~\ref{masses} shows that 
there is {\it some} transition since in the 
normal phase $m_\gamma=0$ within errors. The critical 
region is shown in a magnified form in Fig.~\ref{zoommasses}, 
and one can see that $m_\gamma$ could go to zero continuously
(on the other hand, it should be noted that a discontinuity
in $m_\gamma$ cannot be excluded either).
The dashed curve in Fig.~\ref{masses}
shows the fit $m_\gamma/e_3^2 = A (y_c-y)^{\nu}$
over the whole $y$-range, 
where $A=0.95(2)$, $y_c=-0.046(6)$, 
$\nu=0.44(2)$. For the lower solid curve
$\nu$ has been fixed to $\nu=\fr12$.
In Fig.~\ref{zoommasses} for a smaller $y$-range, $\nu = 0.39(17)$. 
The behaviour of $m_\gamma$ is thus
consistent with a mean field exponent.

\begin{figure}[t]
\hspace{1cm}
\epsfysize=9cm
\centerline{\epsffile{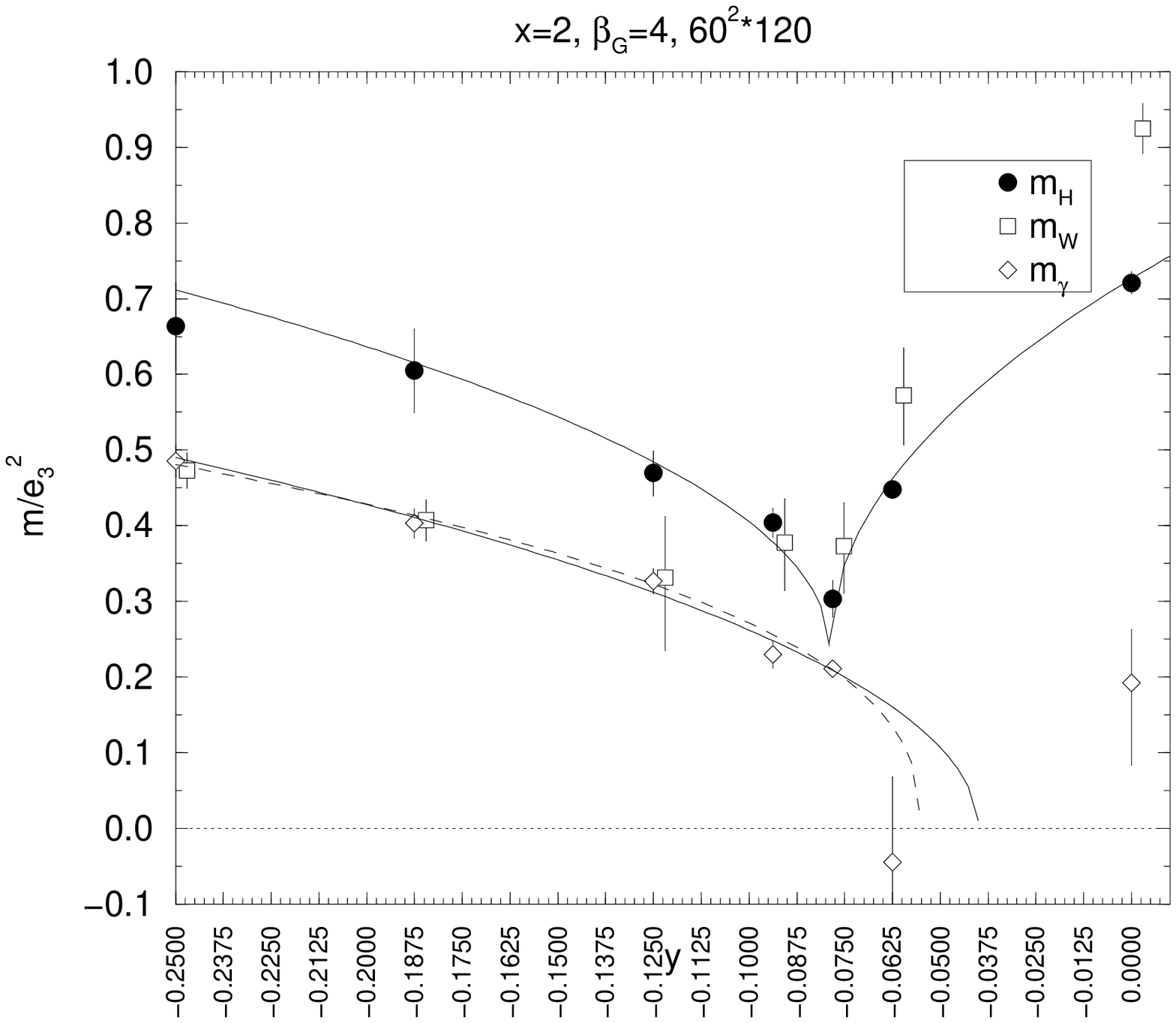}}
\caption[a]{A magnification of the region around
the critical point for $x=2$ in Fig.~\ref{masses}.
The dashed curve is a fit with a free exponent
over the range $y=-0.25\ldots 0.0$, 
and gives $\nu = 0.39(17)$. 
The solid curves are with $\nu=\fr12$, 
and with a constant for $m_H$ (see Fig.~\ref{mHvol}).}
\la{zoommasses}
\end{figure}

As for $m_W$, it is seen that it deviates from $m_\gamma$
already below the critical point, unlike in the 
first order case shown in the top panel of Fig.~\ref{masses}. 
This might indicate that the transition line has split into 
several transitions in the type II regime. 
However, $m_W$ is an excited state and thus there are some
systematic errors in its determination which may be larger
than the statistical errors shown in the figure. 

Consider then the scalar mass $m_H$. First, note
that its minimum is at a point different from 
where $m_\gamma$ diverges (see Fig.~\ref{zoommasses}). Second, $m_H$
dips steeply in Fig.~\ref{masses}, but does not go to zero
as can be seen in Fig.~\ref{zoommasses}. 
On a finite (periodic) lattice with spatial extension 
$L= N a$ one cannot expect to see scalar mass values smaller
than $\sim 2/L$. However, according to Fig.~\ref{mHvol} 
the scalar mass is larger than this value and does
not show the corresponding volume dependence. In fact, $m_H$
is volume-independent within error bars at the minimum. 

The question remaining then is whether
the scalar mass depends on the
lattice spacing $a$. This seems unlikely, since the lattice spacing
we used is much smaller than the correlation length, 
$e_3^2 a=1/\beta_G=1/4 << e_3^2\xi_H=e_3^2/m_H\approx 2$, 
so that one does not expect large effects from removing the 
UV cutoff. Indeed, we have made simulations with $\beta_G=3$
and $\beta_G=6$ at a few points around the minimum, 
and we do not find any appreciable lattice spacing dependence.
It should be pointed out, though, 
that for $\beta_G=6$ one should go to $\sim$50\% larger lattice 
sizes $N$ to get physical volumes comparable with those for $\beta_G=4$,
according to eq.~\nr{constraints}. 

Note also that as long as $a\neq 0$ there
is in a strict sense {\em no} phase transition
due to the Polyakov mass. 
However, the Polyakov mass in eq.~\nr{pmass} is
very small for $\beta_G=4$, so that its finite
value should have no effect. 

\begin{figure}[tbh]
\hspace{1cm}
\epsfysize=9cm
\centerline{\epsffile{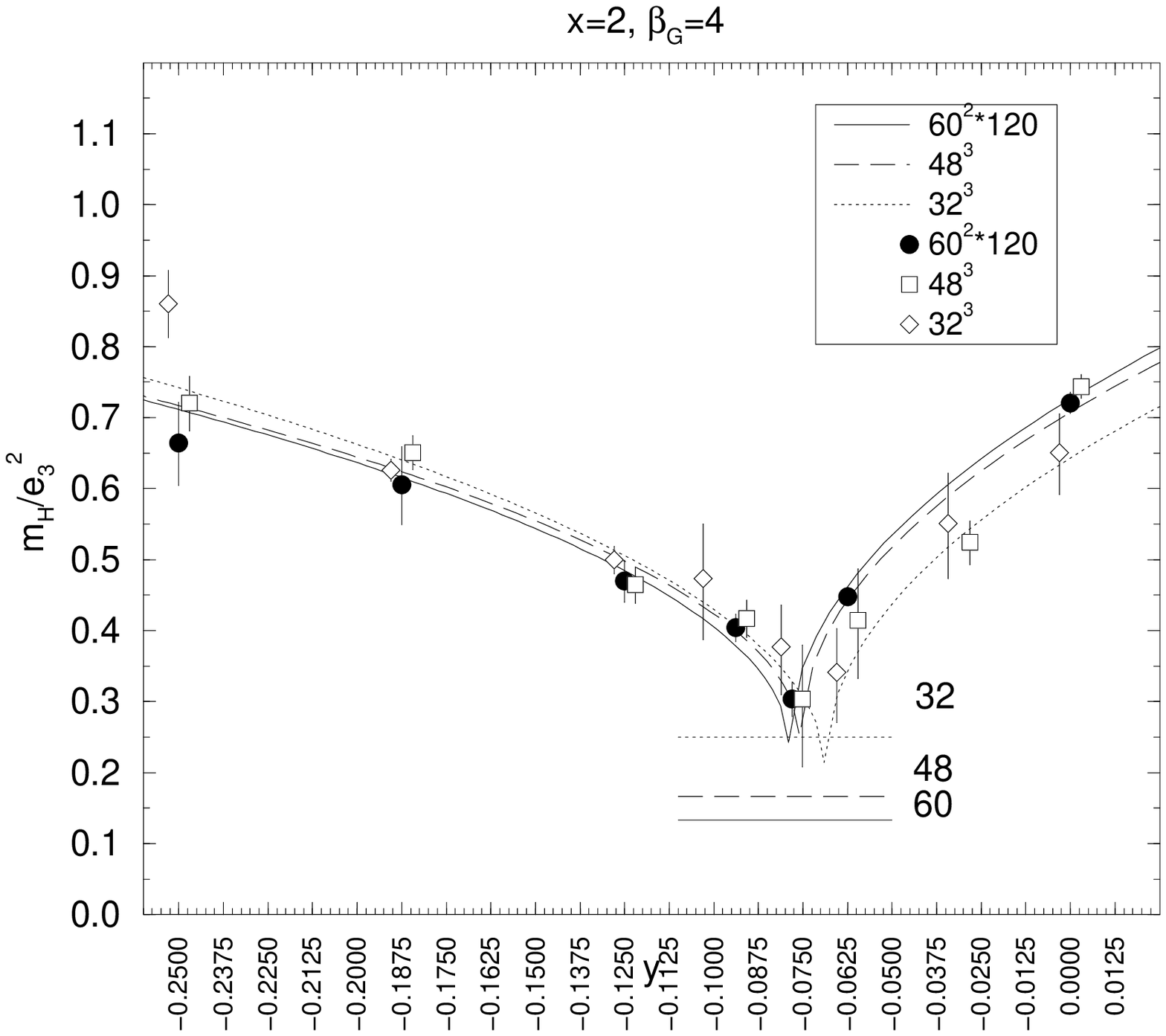}}
\caption[a]{The volume dependence of the scalar mass.
The results are consistent within error bars at the minimum. 
The expected 
critical scaling behaviour is shown with the horizontal 
lines indicating the value of $2/L$. 
No such scaling is observed.
The fits are of the form
$m_H/e_3^2 \sim \left[ A' (y_c-y)^{1/2}+B\right]$ ($y<y_c$),
$\left[A (y-y_c)^{1/2}+B\right]$ ($y>y_c$), with $A'\sim1.1, A\sim1.7$.}
\la{mHvol}
\end{figure}

\section{Discussion}
\la{discussion}

We have seen that in the type II regime, 
the data are consistent with
a second order phase transition driven by a diverging 
photon correlation length.
In contrast, the scalar mass shows quite unexpected
behaviour. It appears that $m_H$ has a minimum away from where
the photon correlation length diverges, and that it does not
show critical scaling towards zero when the volume increases. If this
pattern remains there for larger volumes and smaller lattice spacings, 
then one has to modify the standard picture of the superconducting
phase transition in the type II regime. 

As to the critical exponents of the transition, we
have measured the finite size scaling
susceptibility exponent defined after 
eq.~\nr{susc} to be consistent with zero, and the 
photon correlation length exponent in the SC phase
to be consistent with $\fr12$. We have not measured 
the anomalous dimension $\eta_A$ of the photon correlation
length, $\langle\tilde O_3({\bf x}) \tilde O_3(0)\rangle
\sim 1/|{\bf x}|^{1+\eta_A}$, but in principle it 
could be measured with a significantly extended analysis
(it would show up in the functional
form of eq.~\nr{Gt}). 
If the transition 
were of 2nd order in the usual sense, then one could also
measure the scalar correlation length 
critical exponents $\nu,\nu'$
in the two phases, $m_H \sim |y-y_c|^{\nu,\nu'}$, 
and the corresponding anomalous dimension at $y_c$. However, 
as we have not seen any critical behaviour for the 
scalar correlation length, these exponents cannot be 
systematically measured. It seems that 
away from the critical point, $m_H$ shows approximate
scaling where $\nu,\nu'$ are consistent with $\fr12$, see Fig.~\ref{mHvol}.

A more conclusive study of the infinite volume and continuum
limits in the type II regime would clearly be needed. 
Unfortunately this limit is numerically very demanding and 
requires much more extensive further simulations. 
Based on the present investigation, we can nevertheless point out that 
it would probably be more economic to use the non-compact lattice 
action than the compact one we have used here. The reason is that
one can then take a large lattice spacing 
(smaller $\beta_G$) in order to get a larger physical volume, 
without having to worry about the 
non-zero value of the Polyakov mass.

As a final observation, let us note that one might 
expect the study of a gauge theory with a U(1) gauge 
group to be simpler than that with a more complicated
group such as SU(2)~\cite{isthere?,adjoint} 
or SU(2)$\times$U(1)~\cite{su2u1}. 
In fact, this turns out not to
be the case: U(1) is numerically more demanding 
and requires larger lattices (at least in the compact
formulation). The existence of a massless photon 
is not as such the only reason: there is a photon 
also in SU(2)$\times$U(1), but there
it exists in both phases and is not an order parameter. 

\section*{Acknowledgements}

The simulations, totalling $4\times10^{15} $ flop = 130 Mflops year,
were carried out with a Cray C94 at the Finnish 
Center for Scientific Computing. We acknowledge useful discussions 
with B. Bergerhoff, J. Jers\'ak, S. Khlebnikov, C. Michael,
O. Philipsen, K. Rummukainen, M. Shaposhnikov and G. Volovik.


\begin{references}
\bibitem{othertheories}
It is interesting to note that in some related theories, 
namely an SU(2) gauge theory with a scalar field in the 
fundamental~\cite{isthere?} or adjoint~\cite{adjoint}
representation, there is no first or second order
transition at $x>x_c$.
\bibitem{noncmpct}
If the plaquette part multiplied by $\beta_G$ 
were replaced with the non-compact version, 
then the coefficient 25.5 should be replaced with -1.1.
\bibitem{old}
Indeed, the scalar mass has come down quite a lot since our first
measurements~\cite{klp}. This is because we have now a much better
projection on the ground state and larger physical volumes.
\end{references}
\end{document}